# Telecom-wavelength InAs QDs with low fine structure splitting grown by droplet epitaxy on GaAs(111)A vicinal substrates

A. Tuktamyshev[1, a)], A. Fedorov[2], S. Bietti[1], S. Vichi[1], K.D. Zeuner[3], K.D. Jöns[3], D. Chrastina[4], S. Tsukamoto[1], V. Zwiller[3], M. Gurioli[5], and S. Sanguinetti[1]

[1] L–NESS & Material Science Department, Università degli Studi di Milano Bicocca, via R. Cozzi 55, 20125 Milano, Italy

[2] L-NESS & CNR–IFN, Polo di Como, via F. Anzani 42, 22100 Como, Italy

[3] Department of Applied Physics, Albanova University Centre, KTH Royal Institute of Technology, Roslagstullsbacken 21, 10691 Stockholm, Sweden

[4] L-NESS & Physics Department, Politecnico di Milano, Polo di Como, via F. Anzani 42, 22100 Como, Italy

5)Physics and Astronomy Department, Università degli Studi di Firenze, via G. Sansone 1, 50019 Sesto Fiorentino, Italy

We present self-assembly of InAs/InAlAs quantum dots by the droplet epitaxy technique on vicinal GaAs(111)A substrates. The small miscut angle, while maintaining the symmetries imposed on the quantum dot from the surface, allows a fast growth rate thanks to the presence of preferential nucleation sites at the step edges. A 100nm InAlAs metamorphic layer with In content ≥ 50% directly deposited on the GaAs substrate is already almost fully relaxed with a very flat surface. The quantum dots emit at the 1.3 μm telecom O-band with fine structure splitting as low as 16 μeV, thus making them suitable as photon sources in quantum communication networks using entangled photons.

## I. INTRODUCTION

Entangled photon emitters are fundamental components of the future quantum communication network and the basis of the photonic implementation of quantum information protocols [1,2]. Among possible entangled photon sources, self-assembled quantum dots (QDs) of compound semiconductors are considered as ideal, being able to generate polarization entangled photon pairs on demand via the biexciton (XX)–exciton (X) cascade [1–5]. The presence of the fine structure splitting (FSS) [6,7] of the X state, due to QD anisotropy (shape, composition, etc.), generates a decoherence mechanism, which complicates the observation of the entanglement. Highly symmetric QDs with natural low FSS can be achieved by self-assembled growth on (111) surfaces with $C_{3v}$ symmetry [5,8–10].

The growth of QDs on (111)-oriented surfaces is not straightforward. The common Stranski–Krastanov (SK) growth mode seen in the InAs/GaAs system [11] is not able to induce the self-assembly of QDs on (111) surfaces because of the rapid relaxation of compressive strain due to the low threshold energy for the insertion of misfit dislocations at the substrate-epilayer interface [12,13]. However, by switching the strain from compressive to tensile in the epilayers, self-assembled SK GaAs QDs on InAl(Ga)As(111)A were demonstrated [14–16].

A more efficient and reliable method of obtaining self-assembled QDs on (111) substrates is Droplet Epitaxy (DE) [5,9,10,17,18]. DE relies on kinetically controlled crystallization, via annealing in a group-V atmosphere, of previously formed nanodroplets of group-III metals [19,20]. As DE does not rely on an onset of three-dimensional (3D) growth induced by strain, it is possible to obtain QD self-assembly also in strain-free (e.g., GaAs/AlGaAs) or reduced strain conditions (e.g., InAs/InP). By DE, it is possible to create InAs QDs on GaAs(111)A [21] and InP(111)A [22] substrates, which emit photons at the telecom band for conventional fiber communication. QDs have previously been grown on an InAlAs metamorphic buffer layer (MMBL) on singular GaAs(111)A substrates, using a thin InAs interlayer for complete strain relaxation [21]. The photoluminescence (PL) signals of the QDs indicated a broadband spectrum covering wavelengths from 1.3 to 1.55 μm. The main drawback is the very slow growth rate (below 0.1 ML/s) on singular GaAs(111)A [23].

In our work, we present the DE process on GaAs(111)A for InAs QD emission in the 1.3 μm telecom O-band with an FSS as low as 16 μeV. In order to overcome complications related to the low growth rate on singular (111)A surfaces, we employed a vicinal (111)A surface, where the growth rate is one order of magnitude higher and, thus, is similar to that on GaAs(001) [16,24]. A vicinal surface allows a very flat and almost fully relaxed thin InAlAs MMBL to be obtained, with the root mean square (RMS) roughness of about 1 ML, directly on the GaAs substrate. The measured threading dislocation density (TDD) is of the order of $1\times10^7$ cm$^{-2}$. The TDD was determined using the etch pit density (EPD) approach [25]. This will allow the integration of the self-assembled InAs/InAl(Ga)As QDs into photonic cavity structures (Bragg reflectors) directly on vicinal GaAs(111)A substrates for high brightness efficient entangled photon emitters.

## II. EXPERIMENTAL METHODS

The samples for μ-PL measurements were grown on undoped semi-insulating GaAs(111)A substrates with a miscut of 2° toward $(\bar{1}\bar{1}2)$ in a solid source MBE machine. After a 100 nm GaAs buffer layer grown at 520 °C with a growth rate of 0.5 ML/s, a 200 nm In$_{0.6}$Al$_{0.4}$As barrier layer was deposited at 470 °C with the growth rate of 0.7 ML/s. Then, we supplied indium with the growth rate of 0.01 ML/s at 370 °C, in order to obtain In droplets with a density of about $1\times10^8$ cm$^{-2}$ on InAlAs(111)A. During the indium deposition, the background pressure was below $3\times10^{-9}$ Torr. Then, an As$_4$ flux was supplied for 8 min at the same temperature to crystallize the indium droplets into InAs QDs. After the crystallization process, 10 nm and 140 nm In$_{0.6}$Al$_{0.4}$As capping layers were deposited at 370 °C and 470 °C, respectively, with the growth rate of 0.7 ML/s. The layer structures and growth parameters of the samples are presented in Table I. The morphological characterization of the samples was performed *ex situ* by atomic force microscopy (AFM) in tapping mode, using tips capable of a lateral resolution of about 2 nm.

| Sample | MMBL | In deposition | Annealing in As atmosphere | Capping Layer |
|---|---|---|---|---|
| A | $In_{0.52}Al_{0.48}As$, 470 °C, 100nm | – | – | – |
| B | $In_{0.52}Al_{0.48}As$, 470 °C, 100nm | – | – | – |
| C | $In_{0.6}Al_{0.4}As$, 470 °C, 100nm | – | – | – |
| D | $In_{0.6}Al_{0.4}As$, 470 °C, 200nm | 370 °C, 1 ML | 370 °C, $5\times10^{-5}$ torr | – |
| E | $In_{0.6}Al_{0.4}As$, 470 °C, 200nm | 370 °C, 2 ML | 370 °C, $5\times10^{-5}$ torr | – |
| F | $In_{0.6}Al_{0.4}As$, 470 °C, 200nm | 370 °C, 1 ML | 370 °C, $5\times10^{-5}$ torr | $In_{0.6}Al_{0.4}As$, 10 and 140 nm at 370 and 470 °C |
| G | $In_{0.6}Al_{0.4}As$, 470 °C, 200nm | 370 °C, 2 ML | 370 °C, $5\times10^{-5}$ torr | $In_{0.6}Al_{0.4}As$, 10 and 140 nm at 370 and 470 °C |

**Table I.** Layer structure and growth parameters of the samples presented in this work.

A PANalytical X'Pert PRO MRD high resolution X-ray diffractometer (HR-XRD) equipped with a hybrid mirror and a 2-bounce Ge(220) monochromator was employed for HR-XRD measurements.

To perform PL measurements, the samples were placed in a closed-cycle cryostat and cooled to 8–10 K. To excite the samples, we used a continuous-wave (cw) HeNe laser (632.8 nm). The laser light was focused on the sample using a microscope objective lens with a numerical aperture of 0.8. The spot diameter was of the order of 1 μm. The luminescence signal was collected by the same objective and passed through a beam splitter (10% transmission and 90% reflection) to a spectrometer that contains an 830 line/mm grating. The luminescence signal was spectrally analyzed using a cooled InGaAs photodiode array. A half-wave plate (HWP) and a linear polarizer were inserted into the optical path before the spectrometer, in order to perform polarization-dependent PL measurements.

## III. RESULTS AND DISCUSSION

In order for QDs to emit at telecom wavelengths, it is necessary to use direct-bandgap semiconductor materials with the bandgap below 0.8 eV. One of the III–V semiconductors that may satisfy such a requirement is InAs with its bandgap of 0.35 eV at 300 K and about 0.42 eV at the temperature of liquid He. Unfortunately, InAs QDs embedded in a GaAs matrix emit at a wavelength of about 1 μm [11,26,27] due to the lattice mismatch between GaAs and InAs (about 7%, which affects the maximum size of coherent islands). Therefore, to shift the emission to longer wavelengths and to improve the crystal quality, it is necessary to adapt the heterostructure composition. One possible approach is to fabricate InAs QDs embedded in InGa(Al)As layers, metamorphically grown on GaAs substrates, to reduce the strain between QD and barrier layer materials [28]. This approach was successfully used to shift the InAs QD emission to the telecom band (1.31–1.55 μm) [21,28–30].

### A. Metamorphic substrate optimization on vicinal GaAs(111)A

The realization of an InAl(Ga)As MMBL with the In composition higher than 50% and with a high crystalline quality and a flat surface on a GaAs(111)A substrate is made more difficult by the actual atomic configuration of the surface and by the presence of steps induced by the substrate miscut.

To obtain a fully relaxed InAl(Ga)As MMBL on a singular GaAs(111)A substrate, Mano et al. [21,25] inserted a thin InAs interlayer between the substrate and the MMBL. This has been observed to induce drastic relaxation, due to the introduction of misfit dislocations at the InAs/GaAs interface during the growth of the thin InAs layer [31,32]. It was found that the optimal thickness of the InAs interlayer is 3–7 ML; otherwise, the crystal quality and/or the surface morphology of the InAl(Ga)As MMBL worsen. As a result, a near strain-free metamorphic InAl(Ga)As layer can be formed on singular GaAs(111)A [21,25]. The samples we grew on miscut GaAs(111)A substrates with the insertion of a thin InAs interlayer between the substrate and the InAlAs MMBL did indeed demonstrate full relaxation of the InAlAs MMBL, as in the case of singular GaAs(111)A substrates. However, we also observed the appearance of large islands, with an average lateral size and height of 602±69 and 17.8±4.9 nm, respectively, and with a density of about $7 \times 10^6$ cm$^{-2}$. Such islands can become nucleation sites for droplets and non-radiative recombination centers during the subsequent growth of the QD active layer.

We have found, instead, that on vicinal GaAs(111)A substrates, a fully relaxed InAlAs MMBL, showing a flat surface (RMS roughness < 0.5 nm) and free of large islands, can be obtained by the direct growth of a thin layer (100 nm thickness) with the desired composition. Figure 1(a) displays an XRD reciprocal space map (RSM) for the (224) asymmetric Bragg reflection of sample A. It shows two diffraction peaks that originate from GaAs and $In_{0.52}Al_{0.48}As$. The position of the peak suggests that the indium content in the InAlAs layer is 52.0±0.4%. The peak position of $In_{0.52}Al_{0.48}As$ on the XRD RSM of sample B is the same as that of sample A.

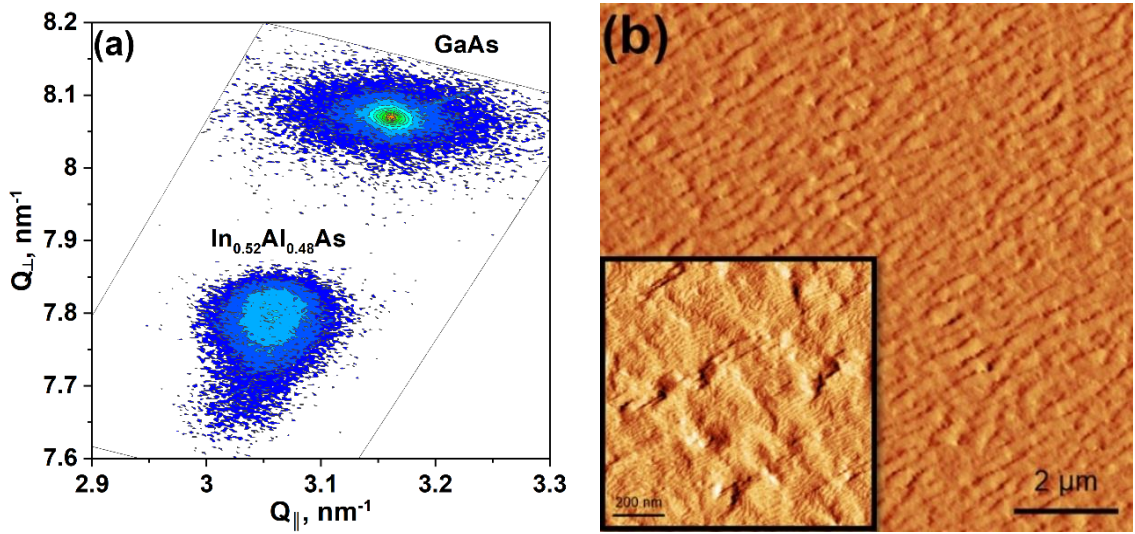

**Figure 1.** (a) XRD reciprocal space map, taken near the (224) asymmetric Bragg reflection of sample A. (b) 10×10 μm² AFM amplitude image of sample C (the inset shows 1×1 μm² AFM tapping amplitude image of the sample).

To find the InAlAs barrier layer content for InAs QD emission at the telecom band, we perform a quantum mechanical 8-band k.p model calculation. The InAs QD has been modeled as a truncated pyramid with a regular triangular base and a small aspect ratio (height to width ratio). As presented below, the actual aspect ratio of our InAs QDs is about 0.05. The InAs QD is surrounded by strain relaxed InAlAs. The simulation suggests using an Al content in the InAlAs layer of less than 50% and a height of the QD of more than 4 nm. Thus, an $In_{0.6}Al_{0.4}As$ barrier layer was chosen for subsequent QD growth. Additionally, the growth of an InAlAs layer with such an In content significantly reduces the strain between the barrier layer and the InAs QDs, which decreases the number of misfit dislocations at the InAs/InAlAs interface improving the optical properties of the QDs. Figure 1(b) shows an AFM image of sample C with 100 nm $In_{0.6}Al_{0.4}As$ directly grown on vicinal GaAs(111)A. The RMS roughness of the surface is observed to be 0.43 nm, calculated from a 1×1 μm² AFM scan, which is comparable to the 1.3 ML thickness of $In_{0.6}Al_{0.4}As$ along [111]. The RMS roughness of sample C, calculated from 5×5 and 10×10 μm² AFM scans, is about 1 nm. Etch pit counting reveals that the TDD is of the order of $1×10^7$ cm$^{-2}$.

**B. Droplet epitaxy quantum dots self-assembly**

In order to study individual QDs by μ-PL, it is necessary to create nanostructures with a density of $10^8 - 10^9$ cm$^{-2}$. From previous work [33], we have determined that In droplets, directly deposited on a vicinal GaAs(111)A substrate at a temperature of about 400 °C, have the desired density.

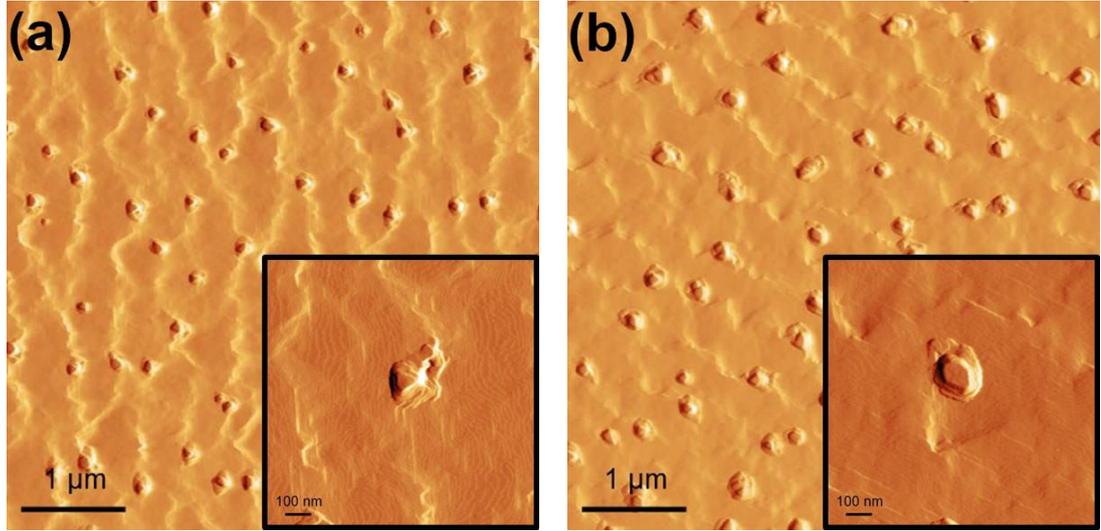

**Figure 2.** (a) 5×5 μm$^2$ AFM tapping amplitude image of sample D (the inset shows 1×1 μm$^2$ AFM tapping amplitude image of an individual QD with triangular pyramidal shape); (b) 5×5 μm$^2$ AFM tapping amplitude image of sample E (the inset shows 1×1 μm$^2$ AFM tapping amplitude image of an individual QD with asymmetrical hexagonal-like pyramidal shape).

Figure 2 shows the morphology of samples D and E with uncapped self-assembled DE InAs QDs fabricated on a 200 nm In$_{0.6}$Al$_{0.4}$As MMBL by deposition of 1 and 2 ML of indium at 370 °C, respectively, followed by annealing in an As$_4$ atmosphere at the same temperature. QD densities for both samples are almost the same: 2.52×10$^8$ cm$^{-2}$ for sample D and 2.50×10$^8$ cm$^{-2}$ for sample E, calculated from 10×10 μm$^2$ AFM scan of each sample. It is worth mentioning that the shape of the QDs is different between samples. Most of the QDs of sample D have a triangular pyramidal shape [see the inset of Fig. 2(a)] with a height of 9.6±2.3 nm and a width of 196±41 nm, measured for 50 QDs. On the other hand, the majority of QDs of sample E have hexagonal-like pyramidal shapes [see the inset of Fig. 2(b)] with a height of 15.9±3.3 nm and a width of 266±52 nm, measured also for 50 QDs. According to Ref. 34, the GaAs DE-QD formation process is strongly affected by the diffusion of Ga adatoms out of the droplet, which leads to the accumulation of GaAs material within a Ga diffusion length of the droplet edge. Using identical crystallization conditions (substrate temperature and As flux), the Ga adatom diffusion length is the same. Considering a model of triangular and hexagonal DE GaAs QD formation on GaAs(111)A [18,35], with the increasing initial droplet size, a shape transition from triangular to hexagonal should occur.

Furthermore, the hexagonal QDs of sample E are elongated in the $\left[1\bar{1}0\right]$ direction along steps due to the presence of a sizeable Ehrlich-Schwöbel (ES) [17], which hinders an adatom diffusivity in the $\left[\bar{1}\bar{1}2\right]$ direction perpendicular to the steps.

**C. Single InAs/InAlAs quantum dot emission**

Samples F and G with capped InAs QDs were characterized by µ-PL. Figure 3(a) shows the broadband PL spectrum of sample F in the range of 800–1500 nm. The peak at 838 nm corresponds to the GaAs substrate, the peaks in the range of 900–1110 nm are associated with defects in the $In_{0.6}Al_{0.4}As$ MMBL (the bandgap of that layer is about 1.32 eV, which corresponds to 940 nm), and the emissions from InAs QDs are placed in the 1100–1350 nm broadband. The broadband PL spectrum of sample G is extremely similar to that of sample F.

A typical PL spectrum of an individual QD of sample F is presented in Fig. 3(b). The observed peak with a linewidth (FWHM) of about 250 µeV (0.33 nm), fitted by a Gaussian function, is attributed to the neutral X line, due to the linear dependence of PL intensity on the excitation power together with the splitting of the emission line into two components, which are linearly polarized in perpendicular directions [36,37]. The observed FWHM of QDs for both samples is in the range of 150—350 µeV. The resolution of the PL setup is 0.07 nm, which corresponds to about 50 µeV at 1310 nm.

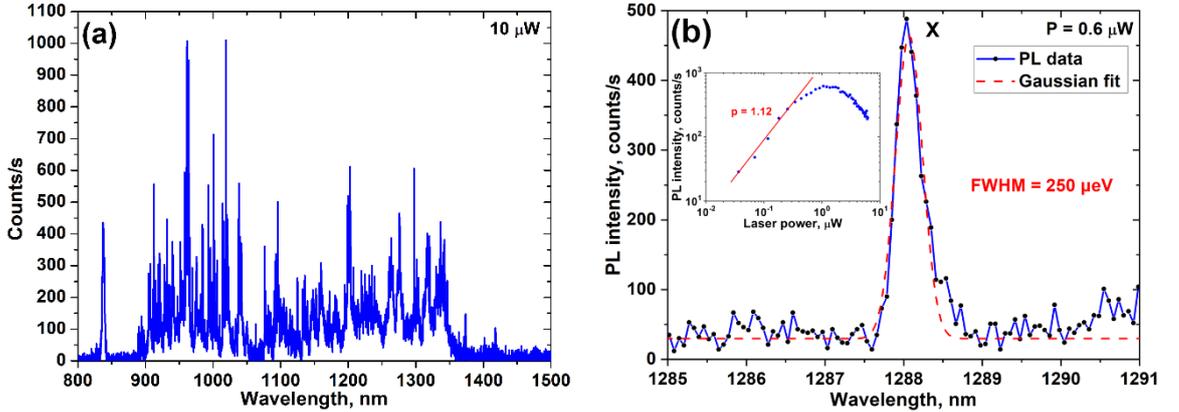

**Figure 3.** (a) The broadband PL spectrum of sample F with a cw excitation of 10 µW; (b) The luminescence spectrum of an individual InAs dot for sample F with a cw excitation of 0.6 µW. The inset shows power dependence of PL intensity of the observed neutral X line.

Polarization-dependent PL measurements were performed for both samples (see Fig. 4). FSS measurements below the limit of the spectrometer were achieved by a polarization sensitive detection method [38,39], which makes it possible to measure FSS with a limit of about 1 µeV using this setup [29]. Figure 4(a) shows a polarization angle dependence of the emission line energy extracted from the center-of-mass PL intensity for an individual QD of sample G, emitting at 1310 nm. The measurement reveals an FSS of 55±4 µeV for this QD.

The most important feature is the existence of telecom QDs with the FSS less than 20 µeV in 20% of the cases [highlighted by a red rectangle in Fig. 4(b)]. The FSS of QDs of sample F ranges from 15 to 100 µeV, with 80% of investigated QDs having the FSS below 60 µeV. As expected, the bigger QDs found on sample G emit at longer wavelengths. Poor statistics for the sample are associated with the fact that most of the observed peaks are related to charged excitons, which do not show FSS, so that just a few neutral lines were observed. The FSS of QDs of sample G is within the range of 55–120 µeV, which is larger than that of the majority of sample F. The larger values of the FSS in sample G in the wavelength region of interest are tentatively associated with the asymmetrical shape of the QDs even if other effects such as the size and strain may play a relevant role.

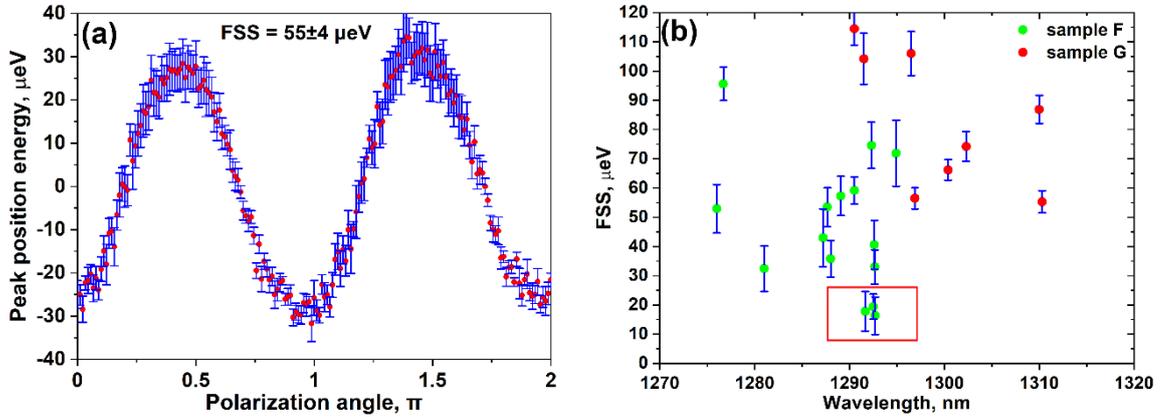

**Figure 4.** (a) Polarization dependence of QD emission line (1310 nm) of sample G with a finite FSS = 55±4 µeV. (b) Statistical distribution of FSS of samples F (green points) and G (red points).

## IV. CONCLUSIONS

In conclusion, we have demonstrated DE growth of InAs QDs embedded in InAlAs layers on vicinal GaAs(111)A substrates, with the high growth rate of a standard (001) substrate process and the high shape symmetry of (111) QDs. A very flat and smooth MMBL with an RMS roughness of about 1 ML has been obtained by direct growth of a 100 nm InAlAs thin layer on GaAs. XRD measurements show that this layer is fully relaxed. The possibility to obtain an MMBL of high quality by using thin epitaxial layers opens the prospect of integrating such structures in cavities realized by distributed Bragg reflectors. In order to perform l-PL characterization of individual QDs, InAs QDs were obtained at a low density of about $2.5 \times 10^8$ cm$^{-2}$. Two types of QD shapes (symmetrical triangular pyramids and hexagonal-like pyramids elongated in the $[1\bar{1}0]$) were observed, depending on the initial size of the droplets.

Fine structure splittings as low as 16±6 μeV at the 1.3 μm telecom O-band suggest the possibility to use these QDs for the future fabrication of entangled photon emitters. The FSS of these dots is similar to that for DE InAs/InP(001) QDs grown by metalorganic vapor phase epitaxy (MOVPE) [40] and slightly higher those of the SK InAs QDs grown on InGaAs/GaAs(001) also by MOVPE [29,30]. To be able to adopt this growth technique for routine use in telecom applications, the samples can be further improved as follows. A rather broad PL linewidth is observed (the mean linewidth is about 250 μeV).We attribute such a behavior to the presence of point defects, due to the low deposition temperature for Al during InAlAs layer growth, the presence of threading dislocations in the InAlAs MMBL, and twin defects after the capping of the InAs QDs. Since InAlAs layers, grown at temperatures above 450 °C, tend to show step bunching with a high number of steps, it is possible to change the MMBL composition from InAlAs to InGaAs. In order to increase the brightness of the dots, InAs QDs could be placed in an InAl(Ga)As cavity between GaAs/Al(Ga)As Bragg reflectors, which can easily be grown on vicinal GaAs(111)A substrates.

## DATA AVAILABILITY

The data that support the findings of this study are available from the corresponding author upon reasonable request.

## ACKNOWLEDGMENTS

The research was supported by the funding from the European Union's Horizon 2020 research and innovation programme under the Marie Skłodowska-Curie grant agreement № 721394.

## REFERENCES

[1] A. Orieux, M. A. M. Versteegh, K. D. Jöns, and S. Ducci, "Semiconductor devices for entangled photon pair generation: a review," *Rep. Prog. Phys.* **80**, 076001 (2017).

[2] D. Huber, M. Reindl, J. Aberl, A. Rastelli, and R. Trotta, "Semiconductor quantum dots as an ideal source of polarization-entangled photon pairs on-demand: a review," *J. Opt.* **20**, 073002 (2018).

[3] J. Skiba-Szymanska, R. M. Stevenson, C. Varnava, M. Felle, J. Huwer, T. Müller, A. J. Bennett, J. P. Lee, I. Farrer, A. B. Krysa, P. Spencer, L. E. Goff, D. Ritchie, J. Heffernan, and A. J. Shields, "Universal growth scheme for quantum dots with low fine-structure splitting at various emission wavelengths," *Phys. Rev. Appl.* **8**, 014013 (2017).

[4] D. Huber, M. Reindl, S. F. C. da Silva, C. Schimpf, J. Martin-Sanchez, H. Huang, G. Piredda, J. Edlinger, A. Rastelli, and R. Trotta, "Strain-tunable GaAs quantum dot: A nearly dephasing-free source of entangled photon pairs on demand," *Phys. Rev. Lett.* **121**, 033902 (2018).

[5] F. Basso Basset, S. Bietti, M. Reindl, L. Esposito, A. Fedorov, D. Huber, A. Rastelli, E. Bonera, R. Trotta, and S. Sanguinetti, "High-yield fabrication of entangled photon emitters for hybrid quantum networking using high-temperature droplet epitaxy," *Nano Lett.* **18**, 505–512 (2018).

[6] D. Gammon, E. S. Snow, B. V. Shanabrook, D. S. Katzer, and D. Park, "Fine structure splitting in the optical spectra of single GaAs quantum dots," *Phys. Rev. Lett.* **76**, 3005–3008 (1996).

[7] M. Bayer, G. Ortner, O. Stern, A. Kuther, A. A. Gorbunovand, A. Forchel, P. Hawrylak, S. Fafard, K. Hinzer, T. L. Reinecke, S. N. Walck, J. P. Reithmaier, F. Klopf, and F. Schafer, "Fine structure of neutral and charged excitons in self-assembled In(Ga)As/(Al)GaAs quantum dots," *Phys. Rev. B* **65**, 195315 (2002).

[8] R. Singh and G. Bester, "Nanowire quantum dots as an ideal source of entangled photon pairs," *Phys. Rev. Lett.* **103**, 063601 (2009).

[9] T. Mano, M. Abbarchi, T. Kuroda, B. McSkimming, A. Ohtake, K. Mitsuishi, and K. Sakoda, "Self-assembly of symmetric GaAs quantum dots on (111)A substrates: Suppression of fine-structure splitting," *Appl. Phys. Express* **3**, 065203 (2010).

[10] T. Kuroda, T. Mano, N. Ha, H. Nakajima, H. Kumano, B. Urbaszek, M. Jo, M. Abbarchi, Y. Sakuma, K. Sakoda, I. Suemune, X. Marie, and T. Amand, "Symmetric quantum dots as efficient sources of highly entangled photons: Violation of Bell's inequality without spectral and temporal filtering," *Phys. Rev. B* **88**, 041306(R) (2013).

[11] J. X. Chen, A. Markus, A. Fiore, U. Oesterle, R. P. Stanley, J. F. Carlin, R. Houdre, M. Ilegems, L. Lazzarini, L. Nasi, M. T. Todaro, E. Piscopiello, R. Cingolani, M. Catalano, J. Katcki, and J. Ratajczak, "Tuning InAs/GaAs quantum dot properties under Stranski-Krastanov growth mode for 1.3 μm applications," *J. Appl. Phys.* **91**, 6710–6716 (2002).

[12] H. Yamaguchi, J. G. Belk, X. M. Zhang, J. L. Sudijono, M. R. Fahy, T. S. Jones, D. W. Pashley, and B. A. Joyce, "Atomic-scale imaging of strain relaxation via misfit dislocations in highly mismatched semiconductor heteroepitaxy: InAs/ GaAs(111)A," *Phys. Rev. B* **55**, 1337–1340 (1997).

[13] H. Wen, Z. M. Wang, J. L. Shultz, B. L. Liang, and G. J. Salamo, "Growth and characterization of InAs epitaxial layer on GaAs(111)B," *Phys. Rev. B* **70**, 205307 (2004).

[14] C. D. Yerino, P. J. Simmonds, B. Liang, D. Jung, C. Schneider, S. Unsleber, M. Vo, D. L. Huffaker, S. Hofling, M. Kamp, and M. L. Lee, "Strain-driven growth of GaAs(111) quantum dots with low fine structure splitting," *Appl. Phys. Lett.* **105**, 251901 (2014).

**[15]** C. F. Schuck, R. A. McCown, A. Hush, A. Mello, S. Roy, J. W. Spinuzzi, B. Liang, D. L. Huffaker, and P. J. Simmonds, "Self-assembly of (111)-oriented tensile-strained quantum dots by molecular beam epitaxy," *J. Vac. Sci. Technol. B* **36**, 031803 (2018).

**[16]** C. F. Schuck, S. K. Roy, T. Garrett, Q. Yuan, Y. Wang, C. I. Cabrera, K. A. Grossklaus, T. E. Vandervelde, B. Liang, and P. J. Simmonds, "Anomalous Stranski-Krastanov growth of (111)-oriented quantum dots with tunable wetting layer thickness," *Sci. Rep.* **9**, 18179 (2019).

**[17]** A. Tuktamyshev, A. Fedorov, S. Bietti, S. Tsukamoto, and S. Sanguinetti, "Temperature activated dimensionality crossover in the nucleation of quantum dots by droplet epitaxy on GaAs(111)A vicinal substrates," *Sci. Rep.* **9**, 14520 (2019).

**[18]** S. Bietti, F. Basso Basset, A. Tuktamyshev, E. Bonera, A. Fedorov, and S. Sanguinetti, "High–temperature droplet epitaxy of symmetric GaAs/AlGaAs quantum dots," *Sci. Rep.* **10**, 6532 (2020).

**[19]** S. Sanguinetti, S. Bietti, and N. Koguchi, "Droplet epitaxy of nanostructures," in Molecular Beam Epitaxy: From Research to Mass Production, 2nd ed. (Elsevier, 2018), Chap. 13, pp. 293–314.

**[20]** M. Gurioli, Z. Wang, A. Rastelli, T. Kuroda, and S. Sanguinetti, "Droplet epitaxy of semiconductor nanostructures for quantum photonic devices," *Nat. Mater.* **18**, 799–810 (2019).

**[21]** N. Ha, T. Mano, T. Kuroda, K. Mitsuishi, A. Ohtake, A. Castellano, S. Sanguinetti, T. Noda, Y. Sakuma, and K. Sakoda, "Droplet epitaxy growth of telecom InAs quantum dots on metamorphic InAlAs/GaAs(111)A," *Jpn. J. Appl. Phys.* **54**, 04DH07 (2015).

**[22]** N. Ha, T. Mano, S. Dubos, T. Kuroda, Y. Sakuma, and K. Sakoda, "Single photon emission from droplet epitaxial quantum dots in the standard telecom window around a wavelength of 1.55 μm," *Appl. Phys. Express* **13**, 025002 (2020).

**[23]** L. Esposito, S. Bietti, A. Fedorov, R. Notzel, and S. Sanguinetti, "Ehrlich-Schwobel effect on the growth dynamics of GaAs(111)A surfaces," *Phys. Rev. Mater.* **1**, 024602 (2017).

**[24]** F. Herzog, M. Bichler, G. Koblmuller, S. Prabhu-Gaunkar, W. Zhou, and M. Grayson, "Optimization of AlAs/AlGaAs quantum well heterostructures on on-axis and misoriented GaAs(111)B," *Appl. Phys. Lett.* **100**, 192106 (2012).

**[25]** T. Mano, K. Mitsuishi, N. Ha, A. Ohtake, A. Castellano, S. Sanguinetti, T. Noda, Y. Sakuma, T. Kuroda, and K. Sakoda, "Growth of metamorphic InGaAs on GaAs(111)A: Counteracting lattice mismatch by inserting a thin InAs interlayer," *Cryst. Growth Des.* **16**, 5412–5417 (2016).

**[26]** W.-H. Chang, W. Y. Chen, T. M. Hsu, N.-T. Yeh, and J.-I. Chyi, "Hole emission processes in InAs/GaAs self-assembled quantum dots," *Phys. Rev. B* **66**, 195337 (2002).

[27] M. Souaf, M. Baira, O. Nasr, M. H. H. Alouane, H. Maaref, L. Sfaxi, and B. Ilahi, "Investigation of the InAs/GaAs quantum dots' size: Dependence on the strain reducing layer's position," *Materials* **8**, 4699–4709 (2015).

[28] L. Seravalli, M. Minelli, P. Frigeri, S. Franchi, G. Guizzetti, M. Patrini, T. Ciabattoni, and M. Geddo, "Quantum dot strain engineering of InAs/InGaAs nanostructures," *J. Appl. Phys.* **101**, 024313 (2007).

[29] M. Paul, F. Olbrich, J. H€oschele, S. Schreier, J. Kettler, S. Portalupi, M. Jetter, and P. Michler, "Single-photon emission at 1.55 µm from MOVPE-grown InAs quantum dots on InGaAs/GaAs metamorphic buffers," *Appl. Phys. Lett.* **111**, 033102 (2017).

[30] K. D. Zeuner, K. D. Jons, L. Schweickert, C. R. Hedlund, C. Nu~nez-Lobato, T. Lettner, K. Wang, S. Gyger, E. Sch€oll, S. Steinhauer, M. Hammar, and V. Zwiller, "On–demand generation of entangled photon pairs in the telecom C–band for fiber–based quantum networks," *arXiv:1912.04782v1* (2019).

[31] A. Ohtake, M. Ozeki, and J. Nakamura, "Strain relaxation in InAs/GaAs(111)A heteroepitaxy," *Phys. Rev. Lett.* **84**, 4665–4668 (2000).

[32] A. Ohtake, T. Mano, and Y. Sakuma, "Strain relaxation in InAs heteroepitaxy on lattice-mismatched substrates," *Sci. Rep.* **10**, 4606 (2020).

[33] A. Tuktamyshev, A. Fedorov, S. Bietti, S. Tsukamoto, R. Bergamaschini, F. Montalenti, and S. Sanguinetti, "Reentrant behavior of the density vs. temperature of indium islands on GaAs(111)A," *Nanomaterials* **10**, 1512 (2020).

[34] S. Bietti, J. Bocquel, S. Adorno, T. Mano, J. G. Keizer, P. M. Koenraad, and S. Sanguinetti, "Precise shape engineering of epitaxial quantum dots by growth kinetics," *Phys. Rev. B* **92**, 075425 (2015).

[35] M. Jo, T. Mano, M. Abbarchi, T. Kuroda, Y. Sakuma, and K. Sakoda, "Self-limiting growth of hexagonal and triangular quantum dots on (111)A," *Cryst. Growth Des.* **12**, 1411–1415 (2012).

[36] S. Kako, C. Santori, K. Hoshino, S. Gozinger, Y. Yamamoto, and Y. Arakawa, "A gallium nitride single-photon source operating at 200 K," *Nat. Mater.* **5**, 887–892 (2006).

[37] M. Abbarchi, C. Mastrandrea, T. Kuroda, T. Mano, A. Vinattieri, K. Sakoda, and M. Gurioli, "Poissonian statistics of excitonic complexes in quantum dots," *J. Appl. Phys.* **106**, 053504 (2009).

[38] K. Kowalik, O. Krebs, A. Lema^ıtre, S. Laurent, P. Senellart, P. Voisin, and J. A. Gaj, "Influence of an in-plane electric field on exciton fine structure in InAs-GaAs self-assembled quantum dots," *Appl. Phys. Lett.* **86**, 041907 (2005).

[39] R. Trotta, E. Zallo, C. Ortix, P. Atkinson, J. D. Plumhof, J. van den Brink, A. Rastelli, and O. G. Schmidt, "Universal recovery of the energy-level degeneracy of bright excitons in InGaAs quantum dots without a structure symmetry," *Phys. Rev. Lett.* **109**, 147401 (2012).

**[40]** T. Muller, J. Skiba-Szymanska, A. B. Krysa, J. Huwer, M. Felle, M. Anderson, R. M. Stevenson, J. Heffernan, D. A. Ritchie, and A. J. Shields, "A quantum light-emitting diode for the standard telecom window around 1550 nm," *Nat. Commun.* **9**, 862 (2018).